\documentclass{ptephy_v1}

\usepackage[latin1]{inputenc}
\usepackage{graphics}
\usepackage{amsmath,amsthm,amssymb}
\usepackage{bm}

\usepackage{amsfonts}
\usepackage{dcolumn}

\usepackage{color}

\newcommand{\beqa}{\begin{eqnarray}}
\newcommand{\eeqa}{\end{eqnarray}}

\begin{document}

\title{Reconstructing $f(R)$ gravity from the spectral index
}

\author{Takeshi Chiba}
\affil{Department of Physics, College of Humanities and Sciences, 
Nihon University, Tokyo 156-8550, Japan}

\begin{abstract}
Recent cosmological observations are in good agreement with 
the scalar spectral index $n_s$ with  $n_s-1\simeq -2/N$, where $N$ is the number of e-foldings. 
In the previous work, the reconstruction 
of the inflaton potential  for a given $n_s$ was studied, 
and  it was found that for $n_s-1=-2/N$, the potential takes the form  
 of either  $\alpha$-attractor model or   chaotic inflation model 
with $\phi^2$ to the leading order in the slow-roll approximation.  
Here we consider  the reconstruction 
of the $f(R)$ gravity model for a given $n_s$ both in the Einstein frame and in the Jordan frame. 
We find that for  $n_s-1=-2/N$ (or more general $n_s-1=-p/N$), $f(R)$ is given parametrically and is found to 
asymptote to $R^2$ for large $R$. This behavior is generic as long as 
the scalar potential is of slow-roll type.  

\end{abstract}
%



\maketitle

\section{Introduction}

The latest Planck data \cite{Planck:2015} are in good agreement with 
the scalar spectral index $n_s$ with  $n_s-1\simeq -2/N$, where $N$ is the number of e-foldings. 
Quadratic chaotic inflation model \cite{chaotic}, Starobinsky model \cite{alex}, Higgs 
inflation with  nonminimal coupling \cite{higgs,higgs2}, and the $\alpha$-attractor  model connecting them with one parameter ``$\alpha$" \cite{kl,kl1,kl2} 
are typical examples  that predict such a relation. 
Are there any other inflation models predicting such a relation?  With this motivation, 
in Ref.\cite{chiba},  we studied such an inverse problem :  
we reconstructed $V(\phi)$ from a given $n_s(N)$ and found that 
for $n_s-1=-2/N$, $V(\phi)$ is either $\tanh^2(\gamma\phi/2)$ (``T-model") \cite{kl,kl1,kl2}
or  $\phi^2$ (chaotic inflation) to the leading order in the slow-roll approximation.  
 
This paper is a continuation of that project: reconstruct $f(R)$ from a given $n_s(N)$.  
Since $f(R)$ gravity in vacuum is equivalent to a scalar field coupled to  Einstein 
gravity via a conformal transformation \cite{tey,whitt,maeda,wands,chiba2} and 
the spectral index is invariant under the conformal transformation \cite{makinosasaki,cy}  
the problem is very simple: convert  
the reconstructed $V(\phi)$  into $f(R)$. We provide such a procedure in Sec.\ref{sec2}. 
We find that as long as the scalar potential is of slow-roll type  $f(R)$ is approximated by $R^2$. 
In fact, for $n_s-1=-2/N$ (or more general $n_s-1=-p/N$), $f(R)$ 
only asymptotes to $R^2$ for large $R$ irrespective of $V(\phi)$.  
In Sec.\ref{sec3}, we also provide the procedure to reconstruct $f(R)$ without relying on the Einstein frame. 
Sec.\ref{secsum} is devoted to the summary.


\section{$f(R) $ from $n_s(N)$: Analysis in the Einstein frame}
\label{sec2}

In this section, we explain the method to reconstruct $f(R)$ 
for a given $n_s(N)$ using the equivalent action in the Einstein frame. 

We study the action  (Jordan frame action) that is given by
\beqa
S=\int \sqrt{-g}d^4x\, \frac{1}{2\kappa^2}f(R) \,,
\label{action:fofr}
\eeqa
where $f(R)$ is a function of the Ricci scalar $R$ and $\kappa^2=8\pi G$.


In order to determine $f(R)$, we utilize the equivalence of $f(R)$ gravity with 
the Einstein-scalar system \cite{tey,whitt,maeda,wands,chiba2}, for which we already have reconstructed 
possible shapes of the potential  
for $n_s-1=-p/N$ \cite{chiba}.  
To show the equivalence, first we note that Eq. (\ref{action:fofr})  
is equivalent to the action 
\beqa
S=\int d^4x\sqrt{-g}\left[\frac{1}{2\kappa^2}f(\psi)+\frac{f_{\psi}}{2\kappa^2}(R-\psi)\right],
\label{action:f(R)2}
\eeqa
where $f_{\psi}=df/d\psi$. 
The variation with respect to the auxiliary field $\psi$ gives $\psi=R$ if $f_{\psi\psi}\neq 0$ 
and the action Eq. (\ref{action:f(R)2}) reduces to $f(R)$ action 
Eq. (\ref{action:fofr}) on-shell. Then,  by the conformal transformation 
 $g_{\mu\nu}=g_{\mu\nu}^E/f_{\psi}$ and by introducing 
 $\kappa\phi= \sqrt{3/2}\ln f_{\psi}$, the action Eq. (\ref{action:f(R)2}) 
can be rewritten as (the so-called  Einstein frame action)
\beqa
S=\int d^4x\sqrt{-g_E}\left[\frac{1}{2\kappa^2}R_E-\frac12 (\nabla_E\phi)^2-V(\phi)\right],
\eeqa
where the quantities with the subscript $E$ denote that defined by 
the metric $g_{\mu\nu}^E$ and $\phi$ and $V(\phi)$ are given using 
$\psi=R$ by
\beqa
\kappa\phi&=& \sqrt{\frac32}\ln f_{R}\equiv  \sqrt{\frac32}\ln F,\\
V(\phi)&=&\frac{FR-f}{2\kappa^2F^2}\,. 
\eeqa
Therefore, once $f(R)$ is given, $V(\phi)$ is determined by the above relation. 
The relation can be converted. Namely, once $V(\phi)$ is given, $f(R)$ is determined by \cite{ms,narain}
\beqa
R&=&e^{\sqrt{2/3}\kappa\phi}\left(\sqrt{6}\kappa V_{,\phi}+4\kappa^2V\right),\label{f(R)fromV1}\\
f(R)&=&e^{2\sqrt{2/3}\kappa\phi}\left(\sqrt{6}\kappa V_{,\phi}+2\kappa^2V\right),
\label{f(R)fromV2}
\eeqa
where $V_{,\phi}=dV/d\phi$. In the following, we set $\kappa=1$. 

We note one important consequence of the relations  (\ref{f(R)fromV1}) 
and  (\ref{f(R)fromV2}). Namely, if $V_{,\phi}$ is negligible compared with $V$, then 
$R\simeq 4e^{\sqrt{2/3}\phi}V=4FV$ and $f\simeq 2F^2V\simeq \frac12 RF$ and hence by 
solving this differential equation it follows that $f$ is approximately proportional to $R^2$ 
\footnote{This explains the results in  Ref.\cite{rinaldi} where it is found numerically that 
in the slow-roll regime $f(R) \propto R^{2-\delta}$ and $\delta$ decreases as 
 the tensor-to-scalar ratio $r$ decreases. In fact,   $r=8(V_{,\phi}/V)^2$.}
 (see Ref.\cite{narain} for a  similar observation).     
Therefore, $R^2$ gravity is quite common in slow-roll inflation. 


\subsection{$n_s-1=-2/N$}

In our previous paper, we  found that $n_s$ can be written in terms of 
the derivative with respect to the e-folding number $N$ as $n_s-1=(\ln (V_{,N}/V^2))_{,N}$ and 
$d\phi/dN=\sqrt{V_{,N}/V}$, where $V_{,N}=dV/dN$. Therefore,  for a given $n_s(N)$, 
we can reconstruct $V(\phi)$. 

In particular, for $n_s-1=-2/N$, the potential is found to be either 
$V(\phi)=\frac12m^2\phi^2$ (chaotic inflation) or 
$V(\phi)=V_0\tanh^2(\gamma\phi/2)$ ($\alpha$-attractor model (T-model)) \cite{chiba}. 
So, using the above relations  (\ref{f(R)fromV1}) and  (\ref{f(R)fromV2}), 
we immediately obtain the corresponding $f(R)$.

\begin{figure}
\centering\includegraphics[height=2in]{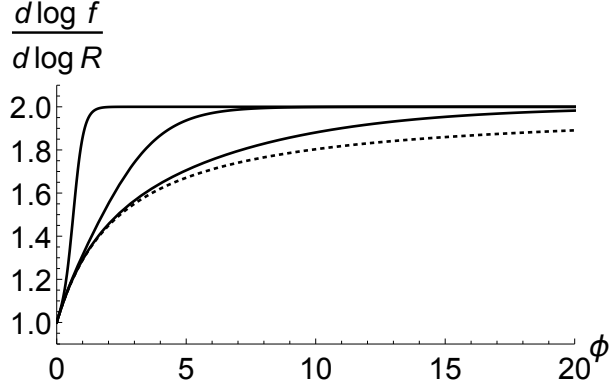}
\caption{\label{fig:index} 
The power index of $f(R)$ as a function of $\phi$. 
The solid curves are for the $\alpha$-attractor model 
with $\gamma=5,\sqrt{2/3},1/5$ from top to bottom. The dotted curve is 
for a quadratic potential. } 
\end{figure}

\subsubsection{Chaotic inflation}

For $V(\phi)=\frac12m^2\phi^2$, from Eq. (\ref{f(R)fromV1}) and Eq. (\ref{f(R)fromV2}), we obtain $f(R)$ parametrically in terms of $\phi$:  
\beqa
&&R=m^2e^{\sqrt{2/3}\phi}\left(\sqrt{6} \phi+2\phi^2\right)=3m^2F(\ln F+(\ln F)^2),
\label{r:chaotic}\\
&&f=m^2e^{2\sqrt{2/3}\phi}\left(\sqrt{6} \phi+\phi^2\right)=
\frac32 m^2F^2(2\ln F+(\ln F)^2).
\label{f:chaotic}
\eeqa
For large $R$,  $R\simeq  2m^2e^{\sqrt{2/3}\phi}\phi^2$ and 
$f\simeq m^2e^{2\sqrt{2/3}\phi}\phi^2 \simeq \frac12 RF$ and hence  $f\propto R^2$,  
as noted above.  For small $R$ (or small $\phi$), 
$f\simeq \sqrt{6}m^2\phi\simeq R$ and the Einstein gravity is recovered. 
In fact, the power index of the functional form of $f(R)$ is calculated by
\beqa
\frac{d\ln f}{d\ln R}=\frac{\sqrt{6}+2\phi}{\sqrt{6}+\phi}.
\eeqa
Hence, since $\phi \sim m^{-1}\sim 10^6$ 
at the beginning of the chaotic inflation, the power index is very close to 2 and 
$f\propto R^2$, and for small $\phi$ the index approaches unity.  Since 
$\phi\simeq 2\sqrt{N}$ for large $N$, the index at the observationally relevant scale 
deviates  from $2$ (see Fig. \ref{fig:index}). 

\subsubsection{$\alpha$-attractor}
For $V(\phi)=V_0\tanh^2(\gamma\phi/2)$, \footnote{Here we have 
fixed an integration constant so that $V(0)=0$. Note that the potential is 
only accurate for large $\gamma\phi$ since the slow-roll approximation is used to reconstruct $V(\phi)$. } 

\beqa
&&R=V_0e^{\sqrt{2/3}\phi}\left(\sqrt{6}\gamma{\rm sech}^2(\gamma \phi/2)\tanh(\gamma\phi/2) +
4\tanh^2(\gamma\phi/2)\right),\\
&&f=V_0e^{2\sqrt{2/3}\phi}\left(\sqrt{6}\gamma{\rm sech}^2(\gamma \phi/2)\tanh(\gamma\phi/2) +
2\tanh^2(\gamma\phi/2)\right) .
\eeqa
For large $R$, 
\beqa
&&R\simeq  V_0e^{\sqrt{2/3}\phi}(4+(4\sqrt{6}\gamma-16)e^{-\gamma\phi})=
4V_0F(1+(\sqrt{6}\gamma-4)F^{-\gamma\sqrt{3/2}}), 
\label{alpha:r}\\
&&f\simeq V_0e^{2\sqrt{2/3}\phi}(2+(4\sqrt{6}\gamma-8 )e^{-\gamma\phi})
=2V_0F^2(1+(2\sqrt{6}\gamma-4 ) F^{-\gamma\sqrt{3/2}}). 
\label{alpha:f}
\eeqa
So, for large $R$, $f\simeq R^2/(8V_0)$ and $R^2$ gravity is approached.   
The Starobinsky model corresponds to 
 $V(\phi)=V_0(1-e^{-\sqrt{2/3}\phi})^2$ and can be approximated by the T-model with $\gamma=\sqrt{2/3}$ for large $\phi$.  \footnote{Setting $\gamma=\sqrt{2/3}$ in Eq. (\ref{alpha:r}) and 
Eq. (\ref{alpha:f}) gives $f\simeq  R^2/(8V_0)+2R+8V_0$. The last two terms are different 
from the Starobinsky model ($f=R^2/(8V_0)+R$), which comes from  neglecting the higher-order terms in  Eq. (\ref{alpha:r}) 
and Eq. (\ref{alpha:f}). }
 The power index of the functional form of $f(R)$ is given by
 \beqa
 \frac{d\ln f}{d\ln R}=\frac{4+(\sqrt{6}\gamma -4)e^{-\gamma\phi}}{
 2+(\sqrt{6}\gamma -2)e^{-\gamma\phi}
}. 
 \eeqa
The index is shown in Fig. \ref{fig:index} for $\gamma=5,\sqrt{2/3},1/5$ 
from top to bottom.

\subsection{$n_s-1=-p/N$}

\begin{figure}
\includegraphics[height=2in]{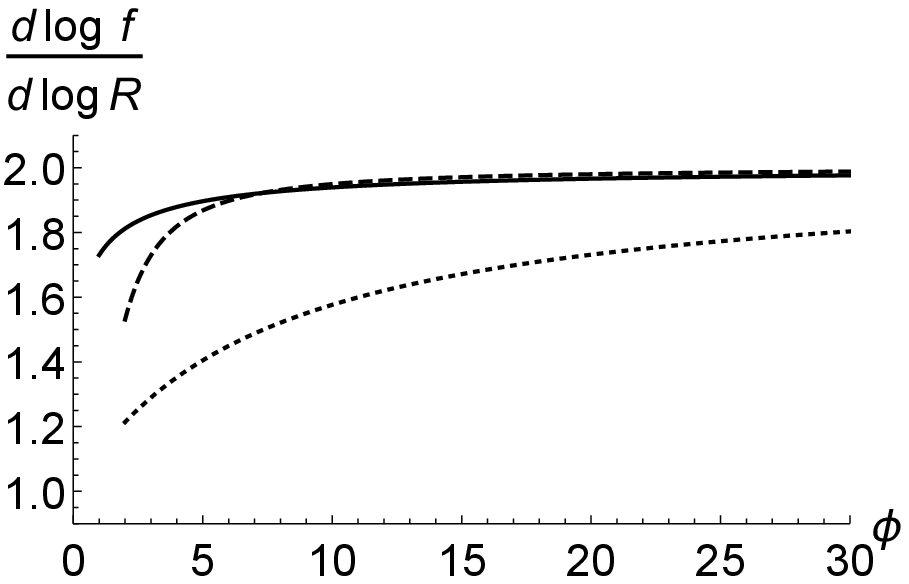}
\includegraphics[height=2in]{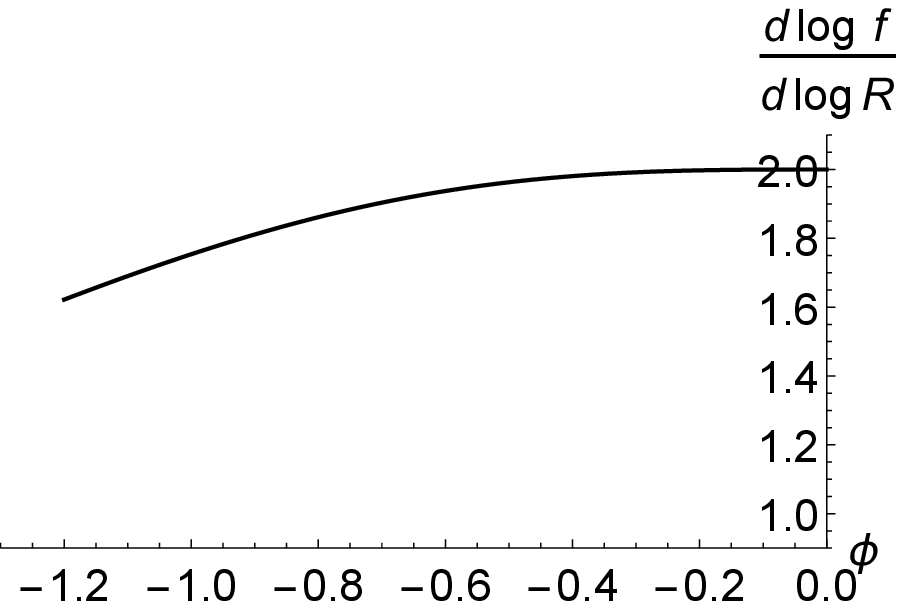}
\caption{\label{fig:indexp} 
The power index of $f(R)$ as a function of $\phi$. 
Left: $p=1/2$ (solid), $p=1$ (logarithmic: dashed), 
$p=4$ (chaotic: dotted). Right: $p=3$ for the first case potential in Eq. (\ref{p/n:potential}). 
} 
\end{figure}

For $n_s-1=-p/N$ ($p\neq 2$ and $p>0$ are assumed), 
the reconstructed potential is \cite{chiba}
\beqa
V(\phi)=
\begin{cases}
V_0-{V_1}\left(\left(1-\frac{p}{2}\right)\phi\right)^{2(p-1)/(p-2)}
  & (p \neq 1)  \\
V_0+V_1\ln \phi & (p=1)\\
 \lambda \phi^{2(p-1)} & (p>1)
 \label{p/n:potential}
\end{cases}
\eeqa
where   in the first case  $\phi\leq 0(\geq 0)$ for $p>2(<2)$. 
  $f(R)$ is constructed using 
Eq. (\ref{f(R)fromV1}) and Eq. (\ref{f(R)fromV2}). In the left-hand panel of 
Fig. \ref{fig:indexp}, the power index of $f(R)$ 
as a function of $\phi$ is shown for $p=1/2$ (solid), $p=1$ (logarithmic: dashed), 
$p=4$ (chaotic: dotted). 
In the right-hand panel, the index for $p=3$ for the first case potential in Eq. (\ref{p/n:potential}) 
is shown. We have assumed $V_0=V_1$. 
We find that in all cases $f(R)$ approaches $R^2$.  

It is interesting to note that although 
in terms of the scalar field potential we have a wide variety of the functional form 
of $V(\phi)$: power-law, exponential, and  logarithmic,  in terms of $f(R)$, for the same $n_s(N)$,  
the functional form is very limited: it only asymptotes to $R^2$ for large $R$.

\section{Analysis in the Jordan frame}
\label{sec3}

Finally, for completeness,  we provide the analysis in the Jordan frame, in which the action is given by Eq. (\ref{action:fofr}). 

The equations of motion in a flat Friedmann universe in vacuum are given by
\beqa
&&F H^2=\frac16(FR-f)-H\dot F,\label{friedmann}\\
&&-2F\dot H=\ddot F-H\dot F,\label{dotH}
\eeqa
where $F=df/dR$ and the dot denotes the derivative with respect to 
the cosmic time $t$, and $R=6(2H^2+\dot H)$. 

We introduce the following slow-roll parameters \cite{Hwang1996,Hwang2001}:
\beqa
\epsilon_1=-\frac{\dot H}{H^2},~~~~\epsilon_2=\frac{\dot F}{2HF},~~~~\epsilon_3=\frac{\ddot F}{H\dot F}\,.
\eeqa
We assume that these parameters are small in the slow-roll approximation. 
In terms of these parameters,  the scalar spectral 
index $n_s$ is given by \cite{Hwang1996,Hwang2001}
\beqa
n_s-1&=&3-2\sqrt{\frac14+\frac{(1+\epsilon_1-\epsilon_2+\epsilon_3)(2-\epsilon_2+\epsilon_3)}{(1-\epsilon_1)^2}}\nonumber\\
&\simeq &-4\epsilon_1+2\epsilon_2-2\epsilon_3-4\epsilon_1^2+2\epsilon_1\epsilon_2-
2\epsilon_1\epsilon_3\,.
\label{ns}
\eeqa
Note that the expression in the first line is exact as long as the slow-roll parameters 
can be regarded as constants. In the second line the expression is expanded 
up to the second order in the slow-roll parameters. 
 We also note that from Eq. (\ref{dotH}) we have 
\beqa
\epsilon_1=-\epsilon_2(1-\epsilon_3). 
\label{epsilon1}
\eeqa
The  e-folding number $N$, which measures the amount of inflationary  
expansion from a particular time $t$ until the end of inflation $t_{\rm end}$, is defined by 
\beqa
N=\int^{t_{\rm end}}_{t}Hdt\,.
\label{efold}
\eeqa
We assume that $N$ is large (say $N\sim O(10)\sim O(10^2)$) under 
the slow-roll approximation.  
In terms of $N$,  from $dN=-Hdt$,  the slow-roll parameters are rewritten as
\beqa
\epsilon_1=\frac{H_N}{H},~~~~\epsilon_2=-\frac{F_N}{2F},~~~~\epsilon_3=-\frac{H_N}{H}-\frac{F_{NN}}{F_N}=-\epsilon_1-\frac{F_{NN}}{F_N}\,,
\label{slow-roll2}
\eeqa
where the subscript $N$ denotes the derivative with respect to $N$. 
Now we introduce a bookkeeping rule to assign  
the order of smallness to the quantities in the slow-roll parameters according to the number of derivatives with respect to $N$: 
\beqa
\frac{F_N}{F}\sim \frac{F_{NN}}{F_{N}}\sim O(1),~~~~~\frac{F_{NN}}{F}\sim O(2)\,.
\eeqa
Then, up to $O(2)$, the spectral index is given by
\beqa
n_s-1=-2\frac{F_N}{F}+2\frac{F_{NN}}{F_N}-\frac32\frac{F_{N}^2}{F^2}.
\label{ns2}
\eeqa

\subsection{$n_s-1=-2/N$}

As an example, let us consider the case of $n_s-1=-2/N$.  The analysis for 
$n_s-1=-p/N$ is similar.

\subsubsection{$O(1)$}
Up to $O(1)$, 
the scalar spectral index in Eq. (\ref{ns2}) is given by
\beqa
n_s-1=-2\frac{F_N}{F}+2\frac{F_{NN}}{F_N}=2\left(\ln F_N/F\right)_N=-\frac{2}{N}
\label{ns:eq}
\eeqa
Then, Eq. (\ref{ns:eq}) is easily integrated to give
\beqa
F(N)=\alpha N^{\beta}, 
\eeqa
where $\alpha$ and $\beta$ are positive constants. \footnote{$\beta>0$ is understood from $\epsilon_1\simeq -\epsilon_2=\beta/(2N)$. }

Then, we use Eq. (\ref{friedmann}) as a differential equation for $f(R)$. 
Since $\dot F=-\beta HN^{\beta-1}=-\beta(F/\alpha)^{(\beta-1)/\beta}$ 
and $R=6H^2(2-\epsilon_1)$, 
Eq. (\ref{friedmann}) becomes up to $O(1)$
\beqa
f=\frac12 RF+\frac18(4-\alpha)\beta R (F/\alpha)^{1-1/\beta} .
\label{f:dalembert}
\eeqa
This type of equation is known as d'Alembert's differential equation, and the general solution is given by
\beqa
R=A\exp\left(\int \frac{\frac12 - \frac{(4-\alpha)(\beta-1)}{8\alpha}(F/\alpha)^{-1/\beta} }{
\frac12 F -\frac{(4-\alpha)\beta}{8}(F/\alpha)^{1-1/\beta} }dF\right),
\eeqa 
where $A$ is an integration constant. 
Up to $O(1)$, the result is
\beqa
R=AF\left(1-\frac{(4-\alpha)(2\beta-1)\beta}{4\alpha}(F/\alpha)^{-1/\beta}\right).
\label{r:sol}
\eeqa
Putting this into Eq. (\ref{f:dalembert}), $f(R)$ is parametrically given by
\beqa
f=\frac12 A F^2\left(1-\frac{(4-\alpha)(\beta-1)\beta}{2\alpha}(F/\alpha)^{-1/\beta}   \right).
\label{f:sol}
\eeqa
We find that by setting $1/\beta \rightarrow \gamma\sqrt{3/2}$ 
these solutions Eq. (\ref{r:sol}) and Eq. (\ref{f:sol}) agree with Eq. (\ref{alpha:r}) 
and Eq. (\ref{alpha:f}). \footnote{Note that $V(\phi)$ in  Eq. (\ref{alpha:r}) 
and Eq. (\ref{alpha:f}) should contain an integration 
constant corresponding to the shift of $\phi$, which has been fixed there. }

\subsubsection{$O(2)$}

However, there can be a case where the $O(1)$ terms in Eq. (\ref{ns2}) 
cancel and only give a higher-order term. For example, let us consider 
$F_N/F=s/N^q$ where $0<q<1$. Then, the $O(1)$ terms in Eq. (\ref{ns2}) are
\beqa
-2\frac{F_N}{F}+2\frac{F_{NN}}{F_N}=-\frac{2q}{N}.
\eeqa
Therefore, we need the $O(2)$ terms to calculate $n_s$. Eq. (\ref{ns2})  up to $O(2)$ becomes 
\beqa
n_s-1=-\frac{2q}{N}-\frac32 \frac{s^2}{N^{2q}}=-\frac{2}{N}. 
\label{ns22}
\eeqa
Since the $O(2)$ term is proportional to $N^{-2q}$, $q=1/2$ is required.\footnote{The corrections due to the time variation of the slow-roll parameters are found to be of even higher order, $O(3)$. }  Then, 
{}from Eq. (\ref{ns22}), we find $s=\sqrt{2/3}$ and $F$ is determined as
\beqa
F(N)=C\exp \left(2\sqrt{\frac{2N}{3}}\right),
\eeqa
where $C$ is a constant. 
Eq. (\ref{friedmann}) becomes up to $O(1)$
\beqa
f=\frac12 RF+\frac12 \frac{RF}{\ln F/C}.
\eeqa
The solution is given by
\beqa
R=A\exp\left(\int  \frac{\frac12+\frac{1}{2\ln F/C}}{\frac{F}{2}-\frac{F}{2\ln F/C}}dF\right) .
\eeqa
Up to $O(1)$, the solution is
\beqa
R=AF(\ln F/C)^2, 
\label{r:sol2}
\eeqa
and $f(R)$ is given by
\beqa
f=\frac12 AF^2(\ln F/C)^2.
\label{f:sol2}
\eeqa
These solutions Eq. (\ref{r:sol2}) and Eq. (\ref{f:sol2}) agree with Eq. (\ref{r:chaotic}) 
and Eq. (\ref{f:chaotic}) for large $R$.

\section{Summary}
\label{secsum}

In this paper, motivated by the relation  $n_s-1\simeq -2/N$ indicated by recent cosmological observations, 
we derived $f(R)$ (the Lagrangian density of $f(R)$ gravity) from $n_s(N)$ in the slow-roll approximation.  
We introduced two approaches to the problem. 
The first approach is to utilize the equivalence of $f(R)$ gravity with the Einstein-scalar system  and 
to determine $f(R)$ from the scalar field potential $V(\phi)$, which is already known \cite{chiba}. 
The second approach is to derive $f(R)$ directly from $n_s(N)$. 
In the first approach, we found that 
 if $V_{,\phi}$ is negligible compared with $V$, then $f(R)$ is approximated by $R^2$. $R^2$ gravity is 
quite common in slow-roll inflation.  

For $n_s-1=-2/N$, we found that $f(R)$ is  determined  parametrically in terms of either $\phi$ 
(Einstein frame case) or $F=f_R$ (Jordan frame case). The results of the two approaches agree. 
The reconstructed $f(R)$ has a common feature: $f(R)\propto R^2$ for 
large $R$.  The results for $n_s-1=-p/N$ are similar.  
In order to recover general relativity at the present time, $f(R)$ is required to satisfy 
$f(R) \simeq R$ at small $R$. 
Therefore, for the same $n_s$, a rather 
restricted functional form of $f(R)$ is allowed, although a wide 
variety of  functional forms of $V(\phi)$ is possible.

\section*{Acknowledgements}

We would like to thank  Masahide Yamaguchi and Nobuyoshi Ohta for useful comments. 
This work is supported by 
MEXT KAKENHI Grant Number 15H05894  and in part
by Nihon University.


\end{document}